# Femtosecond laser ablation by bibursts in the MHz and GHz pulse repetition rates

ANDRIUS ŽEMAITIS,[1,*] MANTAS GAIDYS,[1] PAULIUS GEČYS,[1] MARTYNAS BARKAUSKAS,[2] AND MINDAUGAS GEDVILAS[1]

[1]*Department of Laser Technologies (LTS), Center for Physical Sciences and Technology (FTMC), Savanoriu Ave. 231, 02300 Vilnius, Lithuania*
[2]*Light Conversion Ltd., Keramiku st. 2B, 10233 Vilnius, Lithuania*
*\*andrius.zemaitis@ftmc.lt*

**Abstract:** Here, to the best of our knowledge, for the first time we report the in-depth experimental study of high ultrafast laser ablation efficiency for processing of copper and steel with single-pulses, MHz-, GHz- and burst in the burst (biburst) regime. The comparison of burst, biburst, and single-pulse ablation efficiencies was performed for beam-size-optimised regimes, showing the real advantages and disadvantages of milling and drilling processing approaches. Highly-efficient ultrashort pulse laser processing was achieved for ~1 µm optical wavelength: 8.8 µm$^3$/µJ for copper drilling, 5.6 µm$^3$/µJ for copper milling, and 6.9 µm$^3$/µJ for steel milling. We believe that the huge experimental data collected in this study will serve well for the better understanding of laser burst-matter interaction and theoretical modelling.

## 1. Introduction

To fulfill a high throughput and quality requirements coming from the laser-based manufacturing industry, laser technology must constantly evolve. Therefore, laser source manufacturers build lasers with hundreds of watts of average optical power, pulse repetition rates in the range of MHz and even GHz and near-THz in the burst mode regimes [1,2]. Newly developed fast laser beam scanning systems are capable of reaching scanning speeds of hundreds of meters per second [3]. All the effort is dedicated to make faster laser manufacturing and to keep the laser technology the number one choice for precise material processing. Ultrafast lasers are high-tech products that still hold a high price for know-how and technology, therefore each laser produced photon is very expensive. For example, an ultrafast laser source with an average optical power of tens of watts could easily reach the price of €100k. For this reason, it is extremely important to use laser energy in the most efficient way possible. In the pursuit of higher processing efficiency, the laser sources with burst mode capability were created. These lasers generate packages of pulses called bursts with intra-burst pulse repetition rates up to GHz range. For the conventional, single-pulse laser working regime, it is known that the ablation process indeed can benefit from the high pulse repetition rate as it induces heat accumulation [4]. For the subsequent pulses, the sample is pre-heated due to heat accumulation and the energy required to evaporate the material is lower [5]. In the case of a low repetition rate, the generated thermal energy has enough time between pulses to spread over the sample and surrounding environment. Therefore, heat induced by every previous laser pulse is lost and not beneficial for the ablation process. For GHz burst, the mechanism of material removal via ablation-cooled process was discussed, which claimed ablation efficiency increase due to the removal of excess thermal energy from the material with the successive pulses [6]. Due to high efficiency, the interest in GHz burst laser machining in recent years grew a lot [7–10]. Comparison of single-pulse versus burst mode processing has to be done carefully, since both regimes have to be optimised beforehand and only then compared [11]. In the case of manufacturing processes based on laser ablation such as milling, cutting or drilling, the optimisation of ablation efficiency can be done by varying the laser fluence [12,13]. This optimisation allows finding the most efficient working point, where the highest volume of the material can be removed per unit of energy or/and time. Also, the processing approach has to be taken into account when comparing ablation efficiencies, as different processes might happen during laser drilling and milling as heat accumulation, melt formation, and repulsion of melt out of the processing area.

Here, for the first time we demonstrate an in-depth study of ultrafast laser ablation by bibursts of metals (copper and stainless steel). The biburst laser technology generates burst-in-burst: the package of laser pulses with GHz repetition rate are repeated again at tens of MHz repetition rate burst. The comparison of single-pulse with burst regime in MHz, GHz, and biburst was conducted for the drilling and milling processing approaches. In the case of copper MHz burst processing, a strong influence of odd and even number of pulses within the burst was measured for both drilling and milling. Due to the beam-size-optimisation the highest milling ablation efficiencies were measured of 5.6 $\mu m^3/\mu J$ and 6.9 $\mu m^3/\mu J$ for copper and stainless steel, respectively. At GHz burst processing a big decrease of ablation efficiency was measured for both tested metals. Contrary to GHz burst, the biburst processing demonstrated the high ablation efficiency of 8.4 $\mu m^3/\mu J$ for copper drilling, which was higher than single-pulse drilling efficiency. Contrary, the laser milling approach by bibursts had a lower ablation efficiency than a single-pulse and MHz burst processing modes for both tested metals. In addition, we believe, that the huge experimental data collected in this study will serve well for the better understanding of laser burst-matter interaction and theoretical modelling [14–17].

## 2. Materials and methods

### 2.1 Experimental setup

A solid-state laser (Pharos, Light Conversion) capable of producing light pulses of $\tau = 210$ fs duration at $\lambda = 1030$ nm wavelength was used in the experiments. A galvanometer scanner (Intelliscan 14, Scanlab) together with a F-theta lens with a focal distance of 100 mm was used to scan and focus the laser beam, while the beam waist location and precise sample positioning was performed by linear $Z$- and $XY$-stages (Fig. 1(a)). The state-of-the-art laser had 4 working regimes (Fig. 1 (b)):

1) The conventional single-pulse regime – emitting one pulse every $\Delta t_P = 10$ μs which corresponds to a pulse repetition rate of $f_P = 100$ kHz;

2) MHz burst – emitting burst of pulses with an intra-burst repetition rate of $f_{MHz} = 64.68$ MHz (intra-burst delay time $\Delta t_P = 15.45$ ns) with the number of pulses within the burst ranging from $N = 2$ to $N = 9$ with an inter-burst repetition rate of $f_B = 100$ kHz (inter-burst delay time $\Delta t_B = 10$ μs);

3) GHz burst – emitting burst of pulses with an intra-burst repetition rate of $f_{GHz} = 4.88$ GHz (intra-burst delay time $\Delta t_P = 205$ ps) with the number of pulses within the burst ranging from $P = 2$ to $P = 25$ with an inter-burst repetition rate of $f_B = 100$ kHz (inter-burst delay time $\Delta t_B = 10$ μs);

4) Biburst – tuneable GHz and MHz burst with burst-in-burst regime, where a sequence of $P$ burst pulses at 4.88 GHz is repeated for $N$ times at 64.5 MHz. All combinations of $N = (2, 3, 4, …, 9)$ and $P = (2, 3, 4, …, 25)$ values were possible. In total 192 combinations were theoretically possible for biburst irradiation mode.

The laser was always working at a burst (or biburst) repetition rate of $f_B = 100$ kHz during any of the burst scenarios. The maximum average optical power on the sample surface was $P_{ave} = 7.3$ W, which was always kept constant during the experiments. The maximum pulse energies $E_p$ for single-pulse, MHz burst, GHz burst and biburst regimes were 73 μJ, 36.5 μJ ($N = 2$), 36.5 μJ ($P = 2$), 18.3 μJ ($N = 2, P = 2$), respectively. The shape (envelope) of burst intensity was controlled via laser manufacturers' software. The intensities of intra-burst pulses were set to be as much uniform as possible. Nevertheless, due to the remaining energy in the amplifier, the last pulse in the burst sequence always had a highest intensity (Fig. 2 (c)).

Laser beam radii $w$ along $z$ vertical position were measured by the $D$-squared technique [18]. This technique allows to determine $w$ due to the laser-induced damage diameter $D$ dependence on irradiated pulse energy $E_p$:

$$D^2 = 2w^2 \ln\left(\frac{E_p}{E_{th}}\right), \quad (1)$$

where $E_{th}$ – damage threshold energy. By fitting the experimental data using Equation (1), the beam radii $w$ at different $z$ positions were extracted from the slope of the linear function (Fig. 2 (a)). The Gaussian beam divergence equation was used to fit data obtained by the $D$-squared technique (Fig. 2 (b)) [19]:

$$w(z) = w_0 \sqrt{1 + \left(\frac{(z-z_0)\lambda M^2}{\pi w_0^2}\right)^2}, \quad (2)$$

where $w_0$ – beam radius at waist, $z_0$ – beam waist position, $\lambda = 1030$ nm – laser wavelength, $M^2$ – beam quality factor. The retrieved parameters were: beam radius at focus $w_0 = 19.6 \pm 0.4$ μm and quality factor $M^2 = 1.06 \pm 0.03$. The Rayleigh length was approximately $z_R = 1.1$ mm.

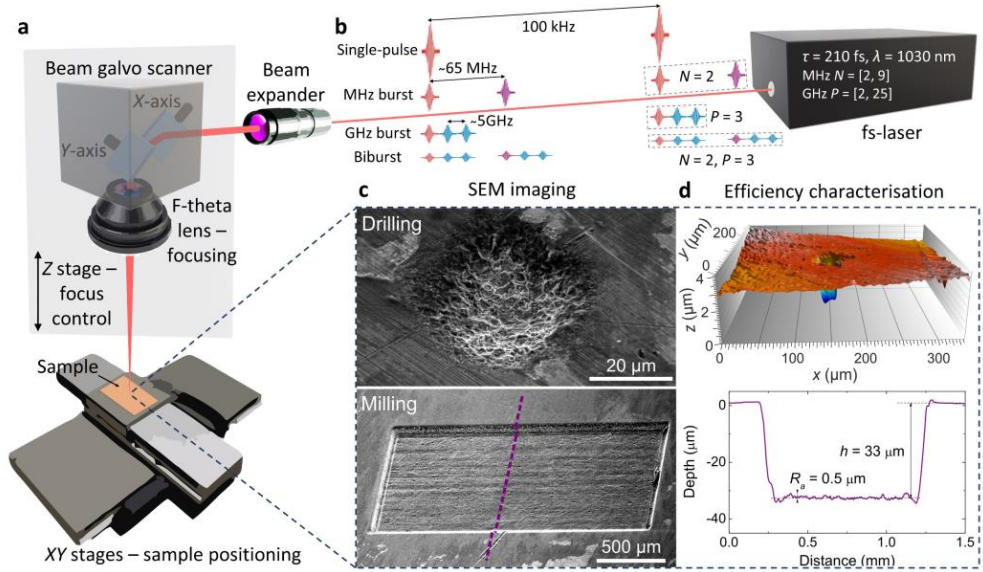

**Fig. 1.** Investigation of ultrafast laser burst mode processing influence on ablation efficiency. (a) Principal scheme of laser processing setup. (b) Illustration of four possible laser working regimes: Single-pulse regime with a pulse repetition rate of $f_P = 100$ kHz; MHz burst with $N = 2$ pulses, an intra-burst repetition rate of $f_{MHz} = 64.68$ MHz and a burst repetition rate of $f_B = 100$ kHz; GHz burst with $P = 3$ pulses, an intra-burst repetition rate of $f_{GHz} = 4.88$ GHz and a burst repetition rate of $f_B = 100$ kHz; Biburst regime with an intra-burst repetition rate of $f_{GHz} = 4.88$ GHz with $P = 3$ pulses within a burst and a burst repetition rate of $f_{MHz} = 64.68$ MHz with $N = 2$ bursts within the biburst and a biburst repetition rate of $f_B = 100$ kHz. (c) SEM images illustrating laser drilling and milling of copper sample and (d) corresponding graphs for efficiency evaluation: map of laser drilled crater measured by the optical 3D profiler, $m = 10$ pulses on one spot; profile of milled cavity measured by the stylus profiler, $n = 3$ scans, beam scanning speed $v = 333$ mm/s, hatch $\Delta y = 10$ μm. Pulse fluence $F_0 = 3.3$ J/cm$^2$, the laser wavelength was $\lambda = 1030$ nm, pulse repetition rate $f_P = 100$ kHz, average optical power $P_{ave} = 7.3$ W.

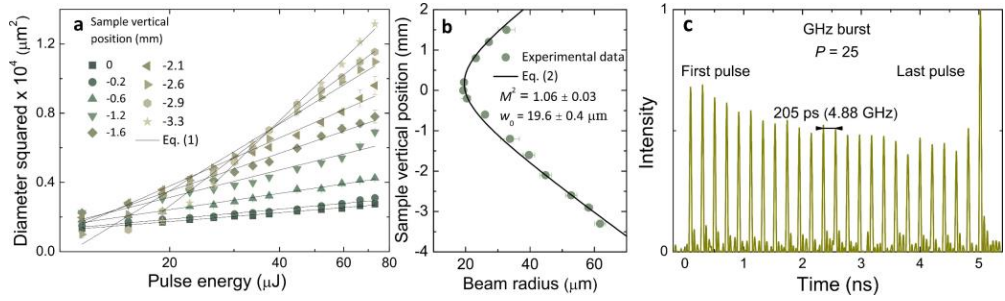

**Fig. 2.** Gaussian beam characterization and GHz burst waveform oscilloscope measurement. (a) Beam radius measurement at different $z$ sample vertical positions by the $D$-squared method. Experimental data fitted by laser damage diameter – pulse energy Eq. (1). (b) Gaussian beam waist calculation by data extracted from the $D$-squared method. Data fitted by the Gaussian beam divergence Eq. (2). (c) Intra-burst intensity distribution of 25-pulse GHz burst.

*2.2 Experimental design*

Laser ablation efficiencies of metal samples were measured by two approaches: 1) ablation of craters – percussion drilling also known as laser punching with fixed beam position, 2) ablation of rectangular cavities – milling by a scanned laser beam. In both approaches for one set of laser processing parameters the maximum ablation efficiency was investigated by changing the beam size from $w = 21$ µm at a position close to $z_0$ focal position to $w = 95$ µm at $\Delta z = 5.3$ mm out of focus. By increasing the distance between the focusing lens and sample surface, the beam size $w$ was increased and therefore the peak pulse fluence $F_0$ was varied:

$$F_0 = \frac{2P_{ave}}{f_{(P,B)} \cdot \pi w^2(z) NP}, \qquad (3)$$

where $P_{ave} = 7.3$ W and $f_{P,B} = 100$ kHz – average optical power and repetition rate which were always kept constant during the experiments, $w(z)$ – beam radius at $z$ position according to Eq. (2), $N$ and $P$ – pulse number within MHz burst and GHz burst, respectively. The ablation efficiencies versus pulse fluence were measured in MHz burst regime for pulses per burst $N = 2, 3, 4, 5, 6, 7, 8, 9$, in GHz burst regime for pulses per burst $P = 2, 3, 4, 5, 6, 7, 8, 9, 10, 15, 20, 25$ and in biburst regime all $N$ and $P$ combinations. Also, the ablation efficiency versus pulse fluence was measured for a single-pulse regime to have a comparison between the conventional single-pulse and burst regimes. In total 1287 (11 beam radii, single-pulse regime, 8 number of pulses per MHz burst, 12 number of pulses per GHz burst and $8 \times 12 = 96$ combinations for biburst) craters and 1287 rectangular cavities were ablated and measured for a copper sample and 1287 rectangular cavities for stainless steel sample.

2.2.1 Laser drilling

Each of the craters was ablated by $m = 10$ bursts on one spot to have a higher depth and volume for more reliable data and a situation closer to the real percussion drilling process by multiple shots. The highest depth of the crater was not higher than 20 µm, which was in linear depth dependence on pulse number on one spot and very far away from crater depth saturation case [20]. Volumes $V$ of the ablated craters were measured by 3D optical profiler (S neox, Sensofar). The ablation efficiency was calculated by dividing the measured volume $V$ by the total accumulated energy on one spot $E_{ACC} = m \cdot P_{ave}/f_{P,B} = 730$ µJ which was always constant during the experiments.

2.2.2 Laser milling

Rectangular cavities with dimensions of 2 mm × 1 mm = 2 mm² were engraved into the metal samples. The rectangles were filled with a pattern of parallel lines separated by $\Delta y = 10$ µm hatch distance. Beam scanning speed of $v = 333$ mm/s was used, resulting in $\Delta x = v/f_B \approx 3.3$ µm spot-to-spot or pitch distance. The rectangles were scanned multiple times $n$ to increase the depths of the cavities for more reliable data and to have a process closer to the industrial laser milling case with multiple layer scan. Typical depths of the cavities were in the range of tens of micrometres. Therefore, the defocusing of the laser beam inside the cavity after the layer scan was negligible since Rayleigh length was close to one millimetre. The depth of the cavity and surface roughness $R_a$ were measured by the stylus profiler Dektak 150+ (Veeko). Ablation efficiency $\eta_E$ for every set of processing parameters was calculated from the cavity depth $h$:

$$\eta_E = \frac{V}{E_{ACC}} = \frac{\Delta y v h}{P_{ave} n}, \qquad (4)$$

where $\Delta y = 10$ µm – hatch distance, $v = 333$ mm/s – beam scanning speed, $P_{ave} = 7.3$ W – average optical power and $n$ – number of scans.

*2.3 Samples*

Copper (CW004A, Ekstremalė) and stainless steel (1.4301, Ekstremalė) plates with dimensions of $50 \times 50 \times 5$ mm$^3$ were used for laser ablation. Copper had a purity of 99.9% and surface roughness of $R_a < 0.1$ µm, while stainless steel surface roughness was $R_a < 0.5$ µm. For sample visualisation, scanning electron microscope (SEM) (JSM-6490LV, JEOL) was used. Copper and stainless steel were chosen as target materials due to high popularity in the theoretical and experimental studies of laser ablation process, which allows easier comparison of the results. Copper was used for laser drilling and milling experiments, stainless steel – for milling.

## 3. Results and discussion

*3.1 MHz burst*

The beam-size-optimisation method allows to simultaneously find the maximum ablation rate and the maximum ablation efficiency for a given set of laser processing parameters [13,21]. This optimisation method was applied for various pulse numbers per burst for drilling of craters and milling of rectangular cavities (see Methods section for the details).

In this paper, all the ablation efficiency versus pulse fluence graphs are either fitted by Equation (S3) and depicted by solid lines or data points connected by straight dashed lines for eye guiding purposes (see the Supplementary material for more information). The equation (S3) does not take into account heat accumulation or plasma shielding which is usually present during high pulse repetition rate processing. Equation (S3) is considered to be for the ideal case of laser processing, where every laser pulse removes the same amount of material without any perturbations. Laser milling due to the multi-pulse (hundreds and thousands of irradiation events per spot size) ablation statistically is more similar to the ideal laser processing than laser drilling (10 irradiation events on one spot). That is why not all experimental data was successfully fitted by Eq. (S3) in Fig. 3 (a) and (b). In some cases of the stainless steel processing (Fig. 4 (a) and (b)) the fitting of Eq. (S3) fails due to the formation of melt-dominated irregular structures finally resulting in stopping of material removal process.

In the case of MHz burst processing of copper, the strong dependence of odd and even number of pulses per burst was clearly visible for both crater ablation and cavity milling (Fig. 3). By using MHz burst the highest ablation efficiency for crater drilling was 8.8 µm$^3$/µJ for $N = 3$ pulses per burst and was higher than single-pulse regime efficiency by 15%. Also, $N = 5$ pulses per burst had a higher crater ablation efficiency by 12.5% compared to single-pulse regime. All other $N$ values were less efficient than the conventional single-pulse regime. For cavity milling, the highest ablation efficiency was 5.6 µm$^3$/µJ for $N = 3$ pulses per burst and was higher than single-pulse regime efficiency by 8%. Similarly, the 3-pulses MHz burst processing was the most efficient regime for pulse-energy-optimisation [22] and beam-size-optimisation [11] for milling, but was never reported for drilling.

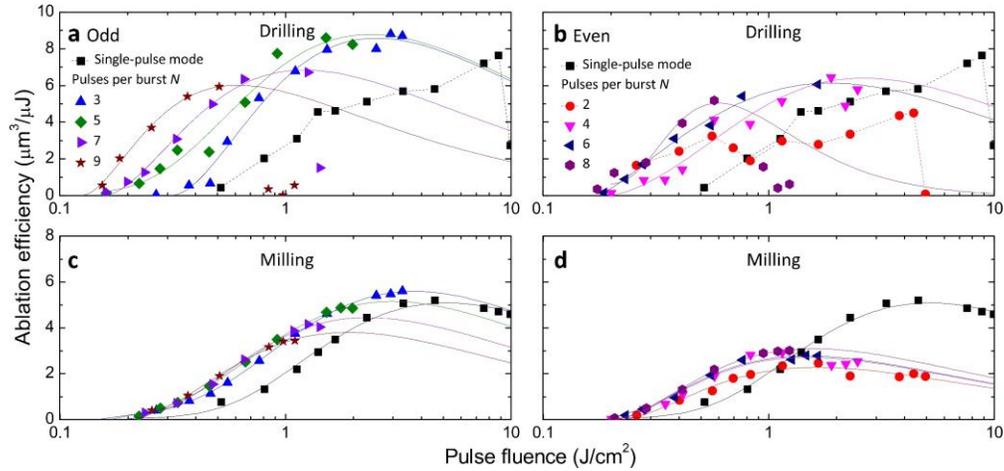

**Fig. 3.** Crater drilling and cavity milling efficiencies by MHz bursts for copper sample. (a, b) Crater drilling, (c, d) cavity milling. (a) and (c) data for odd number of pulses per burst $N$, (b) and (d) – even $N$ values. Black squares are efficiencies for conventional single-pulse laser processing with pulse repetition rate of $f_P$ = 100 kHz. The laser wavelength was $\lambda$ = 1030 nm, burst repetition rate $f_B$ = 100 kHz, intra-burst repetition rate $f_{MHz}$ = 64.68 MHz, average optical power $P_{ave}$ = 7.3 W.

The clear ablation efficiency dependence on number of pulses per burst and processing approach is shown in Fig. 4 (c), where the maximum ablation efficiencies were extracted from Fig. 3. For both crater drilling and cavity milling two laser-initiated processes interchangeably play an important role: one responsible for the reduction of the ablation efficiency at even-pulses burst and the second – for the increase at odd-pulses burst. The process responsible for the reduction of the ablation efficiency is shielding of the second pulse by the plume of ablated particles [23] and plasma [24] produced by the first pulse. Depending on the laser fluence the plume consists of fast atoms and slower nanoparticles [25]. The theoretical studies showed that the depth of the ablated crater by double-pulse configuration is lower than the one for single-pulse ablation [26,27]. This is the consequence of the absorption of the second pulse in the nascent ablation plume, which results in the reheating of ablated material and acceleration of outward part of the plume and deceleration of the inner part of the plume. The similar findings were observed by optical shadowgraphy experiments [23]. The second pulse hits the ablation cloud, the vaporisation of droplets and re-ignition of plasma starts. Due to the second pulse–ablation cloud interaction-induced pressure, part of the material from the ablation cloud might be forced to redeposit back on the target, as a consequence the shielding plume is dispersed [23]. The atomistic simulation of double-pulse ablation confirms the redeposition of material [28]. Also, the measurements by high-precision balances show thrust enhancement for double-pulses with a delay time of 12.2 ns, suggesting the redeposition of material [29].

The third pulse interacts with the target material pre-heated by the redeposited material and also does not suffer the plume attenuation. Therefore, the ablation efficiency is increased as hot material has a higher absorptance [22] and the energy required to raise the temperature to the boiling point of the pre-heated material is lower [30]. The higher volume of the material is ejected by the third pulse, which again creates the ablation plume and all the processes repeat again, resulting in periodical decrease-and-increase in ablation efficiency for the odd and even number of pulses per burst (Fig. 4 (c)). This triangle-wave-type dependency was material and intra-burst repetition rate dependent, since it was measured only for copper at MHz burst and biburst, but not for GHz burst (see later in Fig. 5 (a) and Fig. 7 (a)). Triangle-wave-type dependencies versus number of pulses per burst for copper drilling and milling measured by two completely different processing approaches coincide perfectly, proving that the efficiency measurements are accurate and reliable. The curve of maximum efficiency for drilling was shifted upwards by ~1.5 – 2 times depending on the number of pulses per burst. Similar ~1.5 times more efficient drilling than milling was registered in ref [9]. During burst processing part of the irradiated matter is melted [31]. In the case of crater ablation, some of the matter is removed in liquid form and is seen as burr around the hole. In addition, it is commonly known that after fs laser ablation the condensed particles are found on the workpiece around the processed area as debris. In the case of milling, scanned laser beam interacts multiple time with this burr and debris, therefore energy is consumed to re-heat and evaporate the previously effected matter, that is why the measured efficiency of crater drilling is higher than efficiency of milling [9].

For the steel sample the influence of odd-pulses and even-pulses bursts on ablation efficiency was not observed (Fig. 4 (a), (b)). The highest ablation efficiency was measured for the single-pulse processing mode and was 6.9 $\mu m^3/\mu J$. The MHz burst was more efficient than the single-pulse regime only for pulse fluence values higher than ~2 $J/cm^2$. This can be explained by dense plasma/particle generation at high fluencies. Also, at higher fluence the beam spot size is smaller, therefore the shielding effect is stronger. Therefore, the ablation efficiency close to ~0 $\mu m^3/\mu J$ was measured for single-pulse mode and pulse fluence near ~10 $J/cm^2$. The maximum ablation efficiency of steel dropped down by 35% for 2-pulses burst and 56% for 3-pulses burst compared with single-pulse processing (Fig. 4 (d)). For 4-pulses burst efficiency increased and stabilised at 5-pulses burst, but was still about 28% lower than single-pulse efficiency.

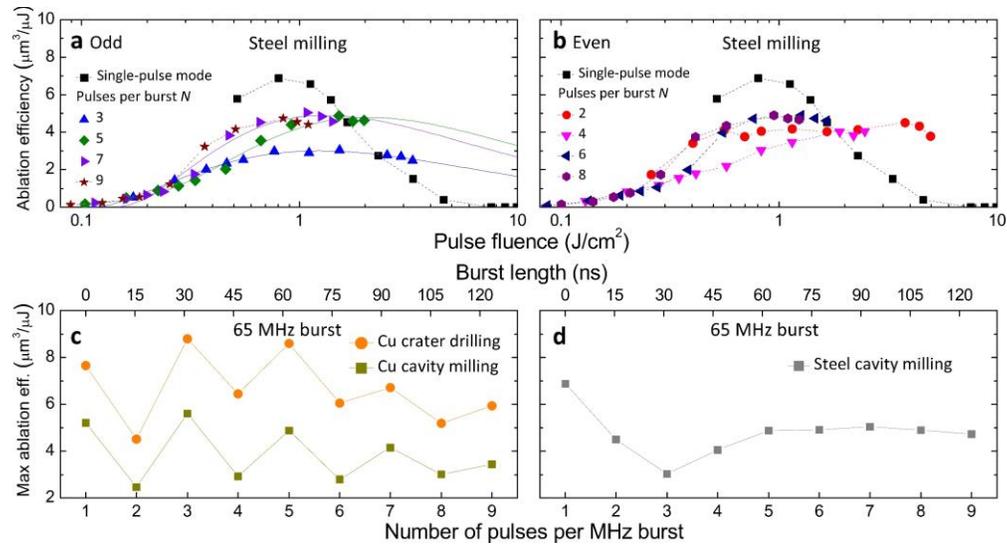

**Fig. 4.** Cavity milling efficiencies of MHz bursts for steel sample for (a) odd and (b) even number of pulses per MHz burst. Black squares are efficiencies for conventional single-pulse laser processing with pulse repetition rate of $f_P$ = 100 kHz. (c) Maximum ablation efficiencies extracted from Fig. 3 versus number of pulses per MHz burst for drilling and milling of copper. (d) Maximum ablation efficiencies extracted from (a) and (b) versus number of pulses per MHz burst for milling of stainless steel. The laser wavelength was $\lambda$ = 1030 nm, burst repetition rate $f_B$ = 100 kHz, intra-burst repetition rate $f_{MHz}$ = 64.68 MHz, average optical power $P_{ave}$ = 7.3 W.

## 3.2 GHz burst

The beam-size-optimisation was applied for GHz burst processing (Fig. 5). The ablation efficiency for copper milling decreased by 78% for $P$ = 2 pulses burst and ~90% for $P$ = 3 and up to $P$ = 25 number of pulses per burst compared to single-pulse milling (5.2 $\mu m^3/\mu J$) regime (Fig. 5 (a)). The similar ablation efficiency decrease was measured for steel milling: for $P$ = 2-pulses burst efficiency decreased by 78%, for $P$ = 3 and more pulses per burst – by 88% - 94% compared to single-pulse milling (6.9 $\mu m^3/\mu J$) regime (Fig. 5 (c)). For GHz burst copper drilling the efficiency was also significantly reduced by 79% - 86% depending on the number of pulses per burst compared to the single-pulse drilling (7.7 $\mu m^3/\mu J$) (Fig. 5 (b)). The maximum efficiency values were extracted from Fig. 5 (a) – (c) and plotted in Fig. 5 (d). The difference between copper drilling and milling was similar to the one measured for MHz burst – depending on the number of pulses per burst drilling was ~1.4 – 2.9 times more efficient than milling. The milling of copper and milling of steel had similar maximum ablation efficiency values versus number of pulses per GHz burst. The high efficiency decrease for GHz burst compared with single-pulse processing was due to the ultrafast laser-matter interaction induced plasma and particle shielding, which partially blocked the incoming laser pulses. In the case of 2-pulses burst processing, 205 ps distance between two pulses was not short enough to prevent attenuation of the second pulse by plasma/particles generated by the first pulse [32].

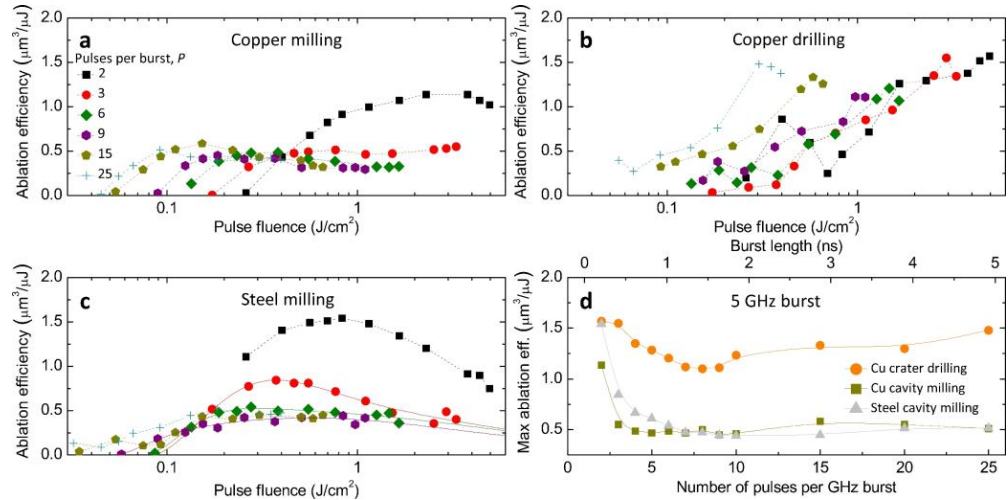

**Fig. 5.** Ablation efficiencies of GHz bursts for (a) copper cavity milling, (b) copper crater drilling and (c) steel cavity milling. (d) Maximum ablation efficiencies extracted from (a), (b), (c) versus number of pulses per MHz burst for milling and drilling of copper and milling of steel. The laser wavelength was $\lambda = 1030$ nm, burst repetition rate $f_B = 100$ kHz, intra-burst repetition rate $f_{GHz} = 4.88$ GHz, average optical power $P_{ave} = 7.3$ W.

*3.3 Biburst*

The beam-size-optimisation was applied for biburst processing. Some of the measurement data are presented in Fig. 6, the rest of the data can be found in the supplementary material. In the case of copper and steel biburst milling the ablation efficiency values were not much different from the GHz burst milling being, at the best, about three times less efficient than single-pulse milling. The unexpected high ablation efficiency values were measured for copper biburst drilling, which at the certain number of pulses per burst combination, exceeded the value of the single-pulse drilling. Nevertheless, the MHz burst drilling was still more efficient than biburst drilling. The difference of biburst drilling and milling efficiencies was huge: for example, for the processing regime $N = 5$, $P = 25$, copper drilling had the efficiency more than 12 times higher than milling (notice the ordinates values in Fig. 6 (a) and (b)). Similarly, the high difference between the efficiencies of milling and drilling of ~10 times was measured for 160-pulse 864 MHz burst and was explained by the different melt flow [16]. In the drilling procedure heat accumulation-induced melt is ejected out of the crater due to the recoil vapour pressure, while during the milling procedure, melt flows back on the previously processed area and does not contribute to the material removal.

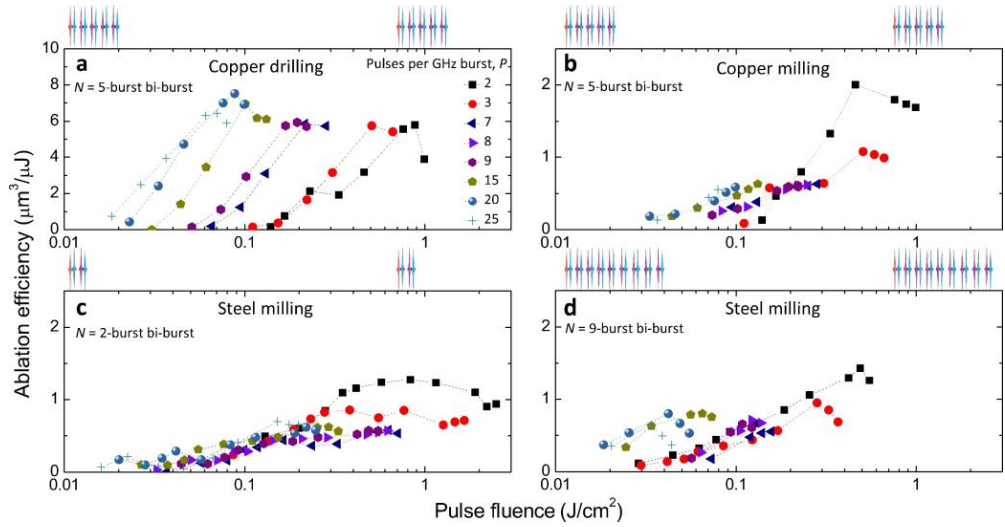

**Fig. 6.** Ablation efficiencies of bibursts for (a) copper crater drilling by $N = 5$ bursts per biburst, (b) copper cavity milling by $N = 5$ bursts per biburst, (c) steel cavity milling by $N = 2$ bursts per biburst and (d) steel cavity milling by $N = 9$ bursts per biburst. Above each of the graphs, the schemes of biburst configuration depicted with $P = 2$ and corresponding $N$. The laser wavelength was $\lambda = 1030$ nm, biburst repetition rate $f_B = 100$ kHz, burst repetition rate $f_{MHz} = 64.68$ MHz, pulse repetition rate $f_{GHz} = 4.88$ GHz, average optical power $P_{ave} = 7.3$ W.

The maximum ablation efficiency values were extracted from all the measured ablation efficiency versus pulse fluence graphs (Fig. 7). In the case of copper milling by bibursts, the influence of odd and even number of bursts per biburst $N$ was evident only for $P = 2$ pulses per GHz burst (Fig. 7 a). This triangle-type-wave dependence was similar to the one measured for the MHz burst processing (Fig. 4 (c)). Contrary, the biburst copper drilling did not have the same shape as MHz burst drilling (Fig. 7 (b)). The drop of ablation efficiency was not measured for $N = 4$ bursts per biburst, ruining the triangle-type-wave graph as was the case for MHz burst processing. For steel biburst milling the small influence of number of bursts per biburst $N$ for maximum ablation efficiency was measured (Fig. 7 (c)). The highest ablation efficiencies achieved in this work for each of the processing mode and approach are summarised in Fig. **7** (d).

SEM images and profiles of the most efficient drilling and milling regimes together with surface roughness are presented in Fig. 8. As shown in our previous works, the ablation efficiency optimisation via beam size [11,21] and pulse energy [33] also results in high-quality. The smallest surface roughness achieved in this work for cavity milling was as low as $R_a = 0.1$ μm showing the promising utilisation of ultrafast bursts in the polishing [34] and high-quality surface treatment [35] applications.

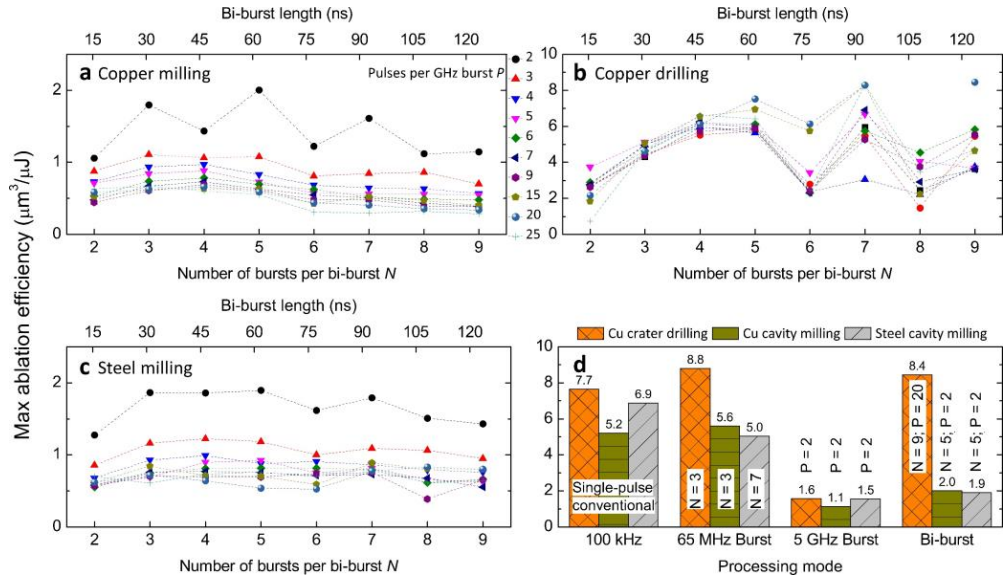

**Fig. 7.** Maximum ablation efficiencies for biburst (a) copper milling, (b) copper drilling and (c) steel milling. (d) The highest ablation efficiencies of drilling and milling measured for four different processing modes. The laser wavelength was $\lambda = 1030$ nm, burst repetition rate $f_{MHz} = 64.68$ MHz, pulse repetition rate $f_{GHz} = 4.88$ GHz, average optical power $P_{ave} = 7.3$ W.

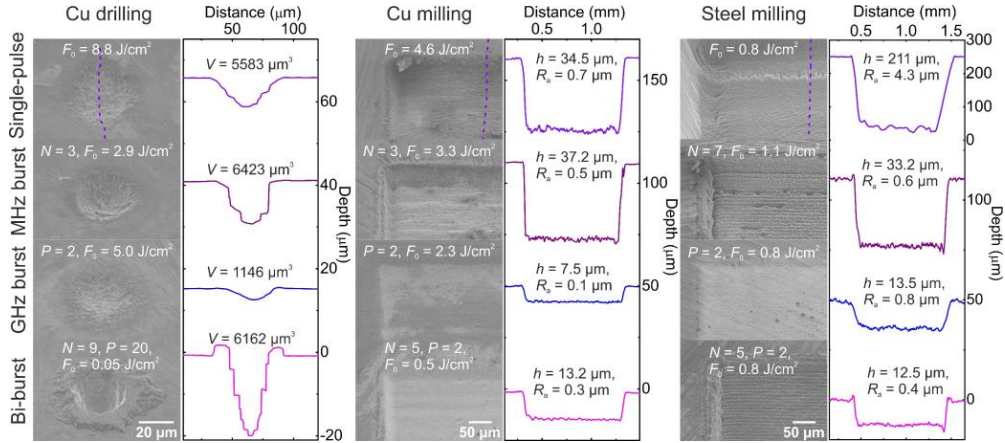

**Fig. 8.** SEM images and profiles of the most efficient drilling and milling regimes for various processing modes. $V$ – volume of the crater, $h$ – depth of the cavity, $R_a$ – surface roughness, $F_0$ – pulse fluence, $N$ and $P$ – number of pulses per MHz and GHz burst, respectively. The laser wavelength was $\lambda = 1030$ nm, repetition rate $f_{P,B} = 100$ kHz, intra-burst repetition rates $f_{MHz} = 64.68$ MHz and $f_{GHz} = 4.88$ GHz, average power $P_{ave} = 7.3$ W.

Overall, the highest ablation efficiency values for copper were measured for MHz burst processing and $N = 3$ pulses per burst and was 8.8 μm$^3$/μJ for drilling and 5.6 μm$^3$/μJ for milling, while the steel milling efficiency was highest for a conventional single-pulse regime with 6.9 μm$^3$/μJ. The biburst processing did not show the highest ablation efficiencies among other processing modes. To the best of our knowledge, in this work, we achieved the highest ever published efficiency values for ultrashort pulses at ~1 μm wavelength. The previous highest ablation efficiency values were 7.6 μm$^3$/μJ for copper drilling [6], 4.8 μm$^3$/μJ for copper milling [11] and 4.1 μm$^3$/μJ for steel milling [31]. For more information about the processing parameters utilised in other studies and typical efficiency, values see Table 1.

**Table 1. Typical ablation efficiency values reported in the literature.** $f_P$ – intra-burst repetition rate, $N$ – number of pulses per burst, $\lambda$ – laser wavelength, $\tau_p$ – pulse duration, $w_0$ – beam radius, $f$ – pulse or burst repetition rate, $v$ – beam scanning speed, $\Delta y$ – hatch distance.

|  |  | Copper |  |  | Stainless steel |  |
|---|---|---|---|---|---|---|
|  | Drilling (µm³/µJ) | $f_P, N, \lambda, \tau_p, w_0, f$ | Milling (µm³/µJ) | $f_P, N, \lambda, \tau_p, w_0, f, v, \Delta y$ | Milling (µm³/µJ) | $f_B, N, \lambda, \tau_p, w_0, f, v, \Delta y$ |
| Burst | 7.6 [6] | 3.456 GHz, 800 ppb, 1035 nm, 1 ps, 12 µm, 1 kHz | 2.6 [36] | 83 MHz, 3 ppb, 1064 nm, 10 ps, 16 µm, 200 kHz, 1.6 m/s, 8 µm | 2.3 [36] | 83 MHz, 3 ppb, 1064 nm, 10 ps, 16 µm, 200 kHz, 1.6 m/s, 8 µm |
| | 6.5 [7] | 1.6 GHz, 400 ppb, 1050 nm, 300 fs, 11.5 µm, 200 kHz | 4.2 [31] | 148 MHz, 28 ppb, 1040 nm, 380 fs, 9 µm, 100 kHz, 750 mm/s, 7.5 µm | 2.5 [31] | 148 MHz, 28 ppb, 1040 nm, 380 fs, 9 µm, 100 kHz, 750 mm/s, 7.5 µm |
| | 8.8 [This work] | 64.68 MHz, 3 ppb, 1030 nm, 210 fs, 23 µm, 100 kHz | 5.6 [This work] | 64.68 MHz, 3 ppb, 1030 nm, 210 fs, 21.7 µm, 100 kHz, 300 mm/s, 10 µm | 5.0 [This work] | 64.68 MHz, 7 ppb, 1030 nm, 210 fs, 24.8 µm, 100 kHz |
| Single-pulse | 0.7 [14] | -, -, 1064 nm, 10 ps, 10 µm, 100 kHz | 2.22 [36] | -, -, 1064 nm, 10 ps, 16 µm, 200 kHz – 1.6 MHz, 8 µm | 2.25 [36] | -, -, 1064 nm, 10 ps, 16 µm, 200 kHz – 1.6 MHz, 8 µm |
| | 1.9 [37] | -, -, 1064 nm, 10 ps, 31.45 µm, 50 Hz | 3.1 [31] | -, -, 1040 nm, 380 fs, 9 µm, 100 kHz, 750 mm/s, 7.5 µm | 4.1 [31] | -, -, 1040 nm, 380 fs, 9 µm, 100 kHz, 750 mm/s, 7.5 µm |
| | 7.7 [This work] | -, -, 1030 nm, 210 fs, 23 µm, 100 kHz | 5.2 [This work] | -, -, 1030 nm, 210 fs, 31.8 µm, 100 kHz, 300 mm/s, 10 µm | 6.9 [This work] | -, -, 1030 nm, 210 fs, 76 µm, 100 kHz, 300 mm/s, 10 µm |

## 4. Conclusions

The in-depth study of maximum ultrafast laser ablation efficiency for processing of copper and steel by single-pulses, MHz-, GHz- and biburst was performed. In the case of copper MHz burst milling and drilling the ablation efficiency was highly dependent on the odd and even number of pulses per burst. The MHz burst drilling was up to two times more efficient than milling process with the same triangle-type-wave dependence on a number of pulses per burst. This type of dependence was material dependent. Steel MHz burst milling had a completely different tendency with no evidence of odd and even number of pulses per burst influence on the ablation efficiency. The GHz processing was revealed to be highly inefficient for both milling and drilling and both copper and steel compared to single-pulse processing. For the first time the biburst mode processing, consisting of GHz bursts inside of MHz bursts, was used for the materials processing. The biburst milling of copper and steel did not improve the ablation efficiency compared to the single-pulse milling. The biburst drilling efficiency of copper had a higher ablation efficiency than the single-pulse drilling. In this paper, we report 3 high efficiency ablation values for ultrashort pulse laser processing at ~1 µm wavelength: 8.8 µm³/µJ (0.5 mm³/min/W) for copper drilling, 5.6 µm³/µJ (0.3 mm³/min/W) for copper milling, and 6.9 µm³/µJ (0.4 mm³/min/W) for steel milling.

**Funding**

Research Council of Lithuania (LMT) (01.2.2-LMT-K-718-03-0050).

**Disclosures**

The authors declare no conflicts of interest.

See Supplement 1 for supporting content.

**References**

1. Y. Liu, J. Wu, X. Wen, W. Lin, W. Wang, X. Guan, T. Qiao, Y. Guo, W. Wang, X. Wei, and Z. Yang, ">100 W GHz femtosecond burst mode all-fiber laser system at 1.0 μm," Opt. Express **28**, 13414–13422 (2020).
2. J. Mur and R. Petkovšek, "Near-THz bursts of pulses – Governing surface ablation mechanisms for laser material processing," Appl. Surf. Sci. **478**, 355–360 (2019).
3. U. Loeschner, J. Schille, A. Streek, T. Knebel, L. Hartwig, R. Hillmann, and C. Endisch, "High-rate laser microprocessing using a polygon scanner system," J. Laser Appl. **27**, S29303 (2015).
4. R. Weber, T. Graf, P. Berger, V. Onuseit, M. Wiedenmann, C. Freitag, and A. Feuer, "Heat accumulation during pulsed laser materials processing," Opt. Express **22**, 11312–11324 (2014).
5. F. Bauer, A. Michalowski, T. Kiedrowski, and S. Nolte, "Heat accumulation in ultra-short pulsed scanning laser ablation of metals," Opt. Express **23**, 1035–1043 (2015).
6. C. Kerse, H. Kalaycıoğlu, P. Elahi, B. Çetin, D. K. Kesim, Ö. Akçaalan, S. Yavaş, M. D. Aşık, Ö. Bülent, H. Heinar, H. Ronald, and F. Ö. Ilday, "Ablation-cooled material removal with ultrafast bursts of pulses," Nature **537**, 84–88 (2016).
7. P. Elahi, Ö. Akçaalan, C. Ertek, K. Eken, F. Ö. Ilday, and H. Kalaycoglu, "High-power Yb-based all-fiber laser delivering 300 fs pulses for high-speed ablation-cooled material removal," Opt. Lett. **43**, 535–538 (2018).
8. K. Mishchik, G. Bonamis, J. Qiao, J. Lopez, E. Audouard, E. Mottay, C. Hönninger, and I. Manek-Hönninger, "High-efficiency femtosecond ablation of silicon with GHz repetition rate laser source," Opt. Lett. **44**, 2193–2196 (2019).
9. G. Bonamis, E. Audouard, C. Hönninger, J. Lopez, K. Mishchik, E. Mottay, and I. Manek-Hönninger, "Systematic study of laser ablation with GHz bursts of femtosecond pulses," Opt. Express **28**, 27702–27714 (2020).
10. M. Gedvilas and G. Račiukaitis, "Spatial zigzag evolution of cracks in moving sapphire initiated by bursts of picosecond laser pulses for ultrafast wafer dicing," RSC Adv. **10**, 33213–33220 (2020).
11. A. Žemaitis, P. Gečys, M. Barkauskas, G. Račiukaitis, and M. Gedvilas, "Highly-efficient laser ablation of copper by bursts of ultrashort tuneable (fs-ps) pulses," Sci. Rep. **9**, 12280 (2019).
12. J. Furmanski, A. M. Rubenchik, M. D. Shirk, and B. C. Stuart, "Deterministic processing of alumina with ultrashort laser pulses," J. Appl. Phys. **102**, 073112 (2007).
13. G. Račiukaitis, M. Brikas, P. Gečys, B. Voisiat, and M. Gedvilas, "Use of High Repetition Rate and High Power Lasers in Microfabrication: How to Keep the Efficiency High?," J. Laser Micro Nanoen. **4**, 186–191 (2009).
14. W. Hu, Y. C. Shin, and G. King, "Modeling of multi-burst mode pico-second laser ablation for improved material removal rate," Appl. Phys. A **98**, 407–415 (2010).
15. M. E. Povarnitsyn, P. R. Levashov, and D. V Knyazev, "Simulation of ultrafast bursts of subpicosecond pulses: In pursuit of efficiency," Appl. Phys. Lett. **112**, 51603 (2018).
16. H. Matsumoto, Z. Lin, and J. Kleinert, "Ultrafast laser ablation of copper with ~GHz bursts," Proc. SPIE **10519**, 1051902 (2018).
17. D. Metzner, P. Lickschat, and S. Weißmantel, "Influence of heat accumulation during laser micromachining of CoCrMo alloy with ultrashort pulses in burst mode," Appl. Phys. A **126**, 84 (2020).
18. J. M. Liu, "Simple technique for measurements of pulsed Gaussian-beam spot sizes," Opt. Lett. **7**, 196–198 (1982).
19. H. Sun, "Thin lens equation for a real laser beam with weak lens aperture truncation," Opt. Eng. **37**, 2906–2913 (1998).
20. A. Žemaitis, M. Gaidys, M. Brikas, P. Gečys, G. Račiukaitis, and M. Gedvilas, "Advanced laser scanning for highly- efficient ablation and ultrafast surface structuring : experiment and model," Sci. Rep. **8**, 17376 (2018).
21. A. Žemaitis, M. Gaidys, P. Gečys, G. Račiukaitis, and M. Gedvilas, "Rapid high-quality 3D micro-machining by optimised efficient ultrashort laser ablation," Opt. Lasers Eng. **114**, 83–89 (2019).
22. B. Jäggi, D. J. Förster, R. Weber, and B. Neuenschwander, "Residual heat during laser ablation of metals with bursts of ultra-short pulses," Adv. Opt. Technol. **7**, 175–182 (2018).
23. D. J. Förster, S. Faas, S. Gröninger, F. Bauer, A. Michalowski, R. Weber, and T. Graf, "Shielding effects and re-deposition of material during processing of metals with bursts of ultra-short laser pulses," Appl. Surf. Sci. **440**, 926–931 (2018).
24. J. König, S. Nolte, and A. Tünnermann, "Plasma evolution during metal ablation with ultrashort laser pulses.," Opt. Express **13**, 10597–10607 (2005).
25. S. Amoruso, R. Bruzzese, C. Pagano, and X. Wang, "Features of plasma plume evolution and material removal

efficiency during femtosecond laser ablation of nickel in high vacuum," Appl. Phys. A **89**, 1017–1024 (2007).
26. M. E. Povarnitsyn, T. E. Itina, K. V Khishchenko, and P. R. Levashov, "Suppression of Ablation in Femtosecond Double-Pulse Experiments," Phys. Rev. Lett. **103**, 195002 (2009).
27. M. E. Povarnitsyn, V. B. Fokin, P. R. Levashov, and T. E. Itina, "Molecular dynamics simulation of subpicosecond double-pulse laser ablation of metals," Phys. Rev. B **92**, 174104 (2015).
28. A. A. Foumani, D. J. Förster, H. Ghorbanfekr, R. Weber, T. Graf, and A. R. Niknam, "Atomistic simulation of ultra-short pulsed laser ablation of metals with single and double pulses: An investigation of the re-deposition phenomenon," Appl. Surf. Sci. **537**, 147775 (2021).
29. D. J. Förster, S. Faas, R. Weber, and T. Graf, "Thrust enhancement and propellant conservation for laser propulsion using ultra-short double pulses," Appl. Surf. Sci. **510**, 145391 (2020).
30. A. Žemaitis, J. Mikšys, M. Gaidys, P. Gečys, and M. Gedvilas, "High-efficiency laser fabrication of drag reducing riblet surfaces on pre-heated Teflon," Mater. Res. Express **6**, 065309 (2019).
31. M. Domke, V. Matylitsky, and S. Stroj, "Surface ablation efficiency and quality of fs lasers in single-pulse mode, fs lasers in burst mode, and ns lasers," Appl. Surf. Sci. **505**, 144594 (2020).
32. J. Schille, L. Schneider, S. Kraft, L. Hartwig, and U. Loeschner, "Experimental study on double-pulse laser ablation of steel upon multiple parallel-polarized ultrashort-pulse irradiations," Appl. Phys. A **122**, 644 (2016).
33. M. Gaidys, A. Žemaitis, P. Gečys, and M. Gedvilas, "Efficient picosecond laser ablation of copper cylinders," Appl. Surf. Sci. **483**, 962–966 (2019).
34. A. Brenner, M. Zecherle, S. Verpoort, K. Schuster, C. Schnitzler, M. Kogel-Hollacher, M. Reisacher, and B. Nohn, "Efficient production of design textures on large-format 3D mold tools," J. Laser Appl. **32**, 12018 (2020).
35. D. Metzner, P. Lickschat, and S. Weißmantel, "High-quality surface treatment using GHz burst mode with tunable ultrashort pulses," Appl. Surf. Sci. **531**, 147270 (2020).
36. T. Kramer, Y. Zhang, S. Remund, B. Jaeggi, A. Michalowski, L. Grad, and B. Neuenschwander, "Increasing the Specific Removal Rate for Ultra Short Pulsed Laser-Micromachining by Using Pulse Bursts," J. Laser Micro Nanoen. **12**, 107–114 (2017).
37. B. Lauer, B. Jaeggi, Y. Zhang, and B. Neuenschwander, "Measurement of the Maximum Specific Removal Rate: Unexpected Influence of the Experimental Method and the Spot Size (M701)," Int. Congr. Appl. Lasers Electro-Optics 146–154 (2015).